# Prediction of 5-year Progression-Free Survival in Advanced Nasopharyngeal Carcinoma with Pretreatment PET/CT using Multi-Modality Deep Learning-based Radiomics

**Bingxin Gu[1†], Mingyuan Meng[2†], Lei Bi[2], Jinman Kim[2], David Dagan Feng[2], Shaoli Song[1,3*]**

[1]Department of Nuclear Medicine, Fudan University Shanghai Cancer Center; Department of Oncology, Shanghai Medical College, Fudan University; Center for Biomedical Imaging, Fudan University; Shanghai Engineering Research Center of Molecular Imaging Probes; Key Laboratory of Nuclear Physics and Ion-beam Application (MOE), Fudan University; Shanghai, PR China.
[2]School of Computer Science, the University of Sydney, Sydney, Australia.
[3]Department of Nuclear Medicine, Shanghai Proton and Heavy Ion Center, Shanghai, PR China.

[†] **Bingxin Gu and Mingyuan Meng contributed equally to this work.**
\* Correspondence: Shaoli Song, shaoli-song@163.com

**Abstract**

**Objective:** Deep Learning-based Radiomics (DLR) has achieved great success in medical image analysis and has been considered as a replacement to conventional radiomics that relies on handcrafted features. In this study, we aimed to explore the capability of DLR for the prediction of 5-year Progression-Free Survival (PFS) in advanced Nasopharyngeal Carcinoma (NPC) using pretreatment PET/CT images.

**Methods:** A total of 257 patients (170/87 patients in internal/external cohorts) with advanced NPC (TNM stage III or IVa) were enrolled. We developed an end-to-end multi-modality DLR model, in which a 3D convolutional neural network was optimized to extract deep features from pretreatment PET/CT images and predict the probability of 5-year PFS. TNM stage, as a high-level clinical feature, could be integrated into our DLR model to further improve the prognostic performance. For a comparison between conventional radiomics and DLR, 1456 handcrafted features were extracted, and optimal conventional radiomics methods were selected from 54 cross-combinations of 6 feature selection methods and 9 classification methods. In addition, risk group stratification was performed with clinical signature, conventional radiomics signature, and DLR signature.

**Results:** Our multi-modality DLR model using both PET and CT achieved higher prognostic performance (AUC = 0.842 ± 0.034 and 0.823 ± 0.012 for the internal and external cohorts) than the optimal conventional radiomics method (AUC = 0.796 ± 0.033 and 0.782 ± 0.012). Furthermore, the multi-modality DLR model outperformed single-modality DLR models using only PET (AUC = 0.818 ± 0.029 and 0.796 ± 0.009) or only CT (AUC = 0.657 ± 0.055 and 0.645 ± 0.021). For risk group stratification, the conventional radiomics signature and DLR signature enabled significant difference between the high- and low-risk patient groups in the both internal and external cohorts ($P < 0.001$), while the clinical signature failed in the external cohort ($P = 0.177$).

**Conclusion:** Our study identified potential prognostic tools for survival prediction in advanced NPC, which suggests that DLR could provide complementary values to the current TNM staging.

# INTRODUCTION

Nasopharyngeal Carcinoma (NPC) is a malignant epithelial cancer. The GLOBOCAN 2020 estimates that there are approximately 133,354 new NPC cases and 80,008 NPC-related deaths worldwide in 2020 (according to the International Agency for Research on Cancer (IARC)) (1). NPC mainly arises from the nasopharynx epithelium, especially the fossa of Rosenmuller (2), and can be pathologically divided into keratinizing differentiated tumor, non-keratinizing differentiated tumor, and non-keratinizing undifferentiated tumor. Due to its unique anatomical structure and radiosensitivity (3, 4), the primary therapeutic regimen for NPC is radiotherapy, with or without chemotherapy, targeted therapy, and immunotherapy. Survival prediction is a major concern in clinical cancer research, as it provides early prognostic information that is needed to guide the therapeutic regimen. In clinical practice, the Tumor, Node, Metastasis (TNM) stage is widely used as an indicator for survival prediction according to the American Joint Committee on Cancer (AJCC)/Union for International Cancer Control (UICC) staging system. However, for patients classified into the same TNM stage, their prognoses may differ widely and their 5-year survival rates range from 10% to 40% for advanced NPC (5-7). This may be attributed to the fact that TNM stage only takes into account of the anatomical information, e.g., size, number, border, and location. Therefore, TNM stage can provide limited benefit for prognoses in patients with advanced NPC (8).

Many clinical biomarkers, such as age, serum lactate dehydrogenase (LDH), body mass index (BMI), and Epstein–Barr virus (EBV), have also been reported as individual prognostic indicators for survival prediction in advanced NPC (9-11). However, these indicators are not specifically relevant to the disease and can be influenced by other indicators, thus failing in repeatability and practicability (12, 13). Non-invasive image-derived biomarkers have also shown good prognostic performance for survival prediction in advanced NPC (14-16). However, conventional imaging modalities, such as Computed Tomography (CT) and Magnetic Resonance Imaging (MRI), only provide tumor's anatomical information. Multi-modality imaging of Positron Emission Tomography / Computed Tomography (PET/CT) provides both anatomical (from CT) and metabolic (from PET) information of the tumor. However, conventional parameters derived from PET/CT, including Standardized Uptake Value (SUV), Metabolic Tumor Volume (MTV), and Total Lesion Glycolysis (TLG), failed to represent intra-tumor information such as tumor texture, intensity, heterogeneity, and morphology (14, 15, 17, 18). Therefore, prognostic indicators that can better represent the tumor characteristics, especially in intra-tumor information, can more accurately predict survival.

Radiomics, as a widely recognized computational method for prognosis, exploits quantitative features (indicators) extracted from medical images to represent tumor characteristics (19). It has drawn much attention among clinical oncologists due to its ability to provide comprehensive representations of tumor characteristics, including intra-tumor information (20). Conventional radiomics refers to the extraction and analysis of high-dimensional handcrafted features from medical images. Through high-throughput feature extraction and statistical machine-learning methods, radiomics can extract and analyze tumor characteristics and has been widely used in many clinical applications (20). Zhang et al. (21) performed radiomics-based prediction of local failure and distant failure in advanced NPC from MRI images. They experimented with 54 cross-combinations derived from 6 feature selection methods and 9 classification methods, and identified optimal combinations in terms of Area Under the receiver-operating characteristic Curve (AUC) and testing error. Lv et al. (22) extracted shape, SUV/HU and texture features from PET and CT images separately and demonstrated these features could improve prognostic prediction for NPC patients. Xu et al. (23) partitioned the primary tumor of NPC from PET/CT images into different subregions, and then separately extracted handcrafted features from these subregions. Yang et al. (24) evaluated the robustness of the handcrafted features extracted from tumor volumes on PET/MR of NPC patients. Conventional radiomics has also been studied in other cancers, such as other head and neck cancers (25) and lung cancer (26). Lv et al.'s research (27) demonstrated multi-level multi-modality fusion radiomics-based model derived from PET/CT outperformed single modality models in prognostic prediction for head and neck cancer patients. These studies demonstrated the capabilities of radiomics for prognosis and identified the optimal conventional radiomics methods for their clinical targets through comprehensive comparisons. However, since conventional radiomics is heavily dependent on human prior knowledge, such as handcrafted feature extraction and manual tuning of many model parameters, its limitations in bringing a source of human bias and lacking the ability to understand high-level semantic information have been well recognized (28, 29).

Recently, deep learning, which leverages deep neural networks to learn deep representations (features) of patterns within images, has achieved great success in medical image analysis and inspired trends toward Deep Learning-based Radiomics (DLR) (30). Compared to conventional radiomics that normally consists of 4 separate steps (**Figure 1A**), DLR adopts a deep neural network to directly predict patient outcomes from the preprocessed image data (**Figure 1B**). DLR can take segmented Regions of Interest (ROIs) as input, but this is optional, with some pipelines that exclude ROI segmentation (30). DLR's feature extraction and feature analysis are jointly learned, in an end-to-end manner, which thereby removes the reliance on time-consuming handcrafted feature extraction, and allows for automatic learning of relevant and robust features without human intervention (28). In other words, DLR can remove the human bias brought by handcrafted features and potentially discover high-level semantic features that may be overlooked by manually-defined feature extraction.

DLR has been widely used in the studies of many cancers, including glioma (31), lung cancer (32), breast cancer (33), renal tumor (34), and NPC (35-38). Peng et al. (35) proposed one of the earliest studies where DLR was introduced into the prognosis of NPC. They used pre-trained Convolutional Neural Networks (CNNs) to extract deep features from PET and CT images separately, and then fed the deep features, as well as conventional handcrafted features, into a Cox proportional hazard regression model to establish a prognostic nomogram. Their study suggested that deep features can serve as reliable and powerful indicators for prognosis, but their feature extraction and analysis were not joint learned in an end-to-end manner. Existing end-to-end DLR studies (31-34, 36-38) are limited by: (1) they were mainly designed for a single imaging modality such as MRI and CT, so their DLR models cannot derive complementary features from multi-modality PET/CT images; and (2) they had limited comparison to the conventional radiomics methods (e.g., only a few conventional radiomics methods were included for comparison), which undermines the reliability of their conclusions.

In this study, we aimed to develop an end-to-end multi-modality DLR model to directly predict 5-year Progression-Free Survival (PFS) from pretreatment PET/CT images, and perform a comprehensive comparison with conventional radiomics methods. Our DLR model is a 3D CNN that was purposely optimized for multi-modality PET/CT images and can simultaneously extract complementary deep features from both PET and CT images. Our DLR model can integrate TNM stage as a high-level clinical feature and this has been demonstrated to further improve prognostic performance.

## MATERIAL AND METHODS

### Patients and PET/CT Image Data

From November 2009 to May 2019 and June 2014 to May 2019, the medical records of 281 NPC patients from Fudan University Shanghai Cancer Center (FUSCC) and 178 NPC patients from Shanghai Proton and Heavy Ion Center (SPHIC) undergoing PET/CT were retrospectively screened, respectively. The eligibility criteria were as follows: (i) TNM stage III or IVa according to the 8th edition of AJCC guidelines, (ii) the follow up time of patients without progression more than 5 years, and (iii) available clinical data and PET/CT imaging data. Finally, 170 patients from FUSCC and 87 patients from SPHIC were enrolled in this study (**Figure 2**). All patients received therapeutic regimens at FUSCC or SPHIC according to the National Comprehensive Cancer Network (NCCN) guidelines. The detailed therapeutic regimen is presented in the **Supplementary Material**.

After completion of initial treatment, physical examination, imaging examination, and nasopharyngoscopy were performed every 3 months in the first 2 years, then every 6 months in the third to fifth year, and once a year thereafter. Treatment responses were identified by imaging examination according to RECIST 1.1 (39). The following endpoint was set as PFS, defined as the time from randomization to disease progression (locoregional or distant) or death (3). The median follow-up time was 89 months (range, 61-149 months) for FUSCC and 65 months (range, 29-92 months) for SPHIC. FUSCC and SPHIC Ethical Committee approved this study, and informed consent was obtained from all enrolled patients.

All patients underwent $^{18}$F-fluorodeoxyglucose ($^{18}$F-FDG) PET/CT (Siemens Biograph 16HR, Knoxville, Tennessee, USA) prior to treatment within 4 weeks. Detailed data acquisition and reconstruction were recorded in the **Supplementary Material**.

**Problem Definition**

We aimed for long-time survival prediction in patients with advanced NPC using pretreatment PET/CT images. In this study, we mainly focused on 5-year PFS and regarded the survival prediction as a binary classification problem. Specifically, the patients with PFS≤60 months (5 years) were labeled as 1, while other patients (PFS>60 months) were labeled as 0. Then, the objective of this study was to classify patients into these two classes (0 or 1) using pretreatment PET/CT images.

**Internal and External Cohorts**

The 170 patients acquired from FUSCC were assigned to an internal cohort, while the 87 patients acquired from SPHIC were assigned to an external cohort. The internal cohort was used for the establishment and internal validation of all prognostic methods. Specifically, each method was trained and validated using 5-fold cross-validation within the internal cohort, and all hyper-parameters were decided based on the internal validation. For external validation, the 5 models of each method, established at the 5-fold cross-validation, were validated on the external cohort. It should be noted that the external cohort is completely independent from the model establishment.

**Data Preprocessing**

All PET/CT images were preprocessed through the following steps: (1) Primary tumors were segmented on PET and CT images simultaneously using a semi-automatic segmentation algorithm, which is available with the ITK-SNAP software (version 3.8.0, http://www.itksnap.org). The semi-automatic segmentation results were then manually adjusted and refined by a senior nuclear medicine physician (SL-S) to ensure reliability. (2) PET images were normalized based on body mass. The derived body mass was applied to convert PET images into SUV maps. (3) PET/CT images and tumor segmentation masks were resampled into isotropic voxels of unit dimension to ensure comparability, where 1 voxel corresponds to 1 mm$^3$. Specifically, PET/CT images and segmentation masks were resampled via linear interpolation and nearest neighbor interpolation, respectively. (4) PET/CT images were multiplied with the binary masks of tumor segmentation to extract ROI, and then were cropped into three-dimensional patches with a size of 80 × 80 × 64 voxels. The patch size was selected to ensure that the whole segmented tumor can be included and the tumor center aligned with the patch center. (5) The patches were then normalized to range [0, 1] individually using the 99th percentile pixel value and 0 as upper and lower limits. The pixels whose values are higher/lower than the upper/lower limit were cut off to be 1/0.

**Conventional Radiomics Analysis**

For performing a comprehensive comparison between DLR and conventional radiomics, we first followed the analytic scheme proposed by Zhang et al. (21) to identify the optimal conventional radiomics methods for our research objective, and then the optimal methods were chosen as benchmarks and compared to our DLR model in the following analysis. Specifically, a total of 6 feature selection methods based on statistical approaches were used in the analysis: $L^1$-Logistic Regression ($L^1$-LOG), $L^1$-Support Vector Machine ($L^1$-SVM), Random Forest (RF), Distance Correlation (DC), Elastic Net Logistic Regression (EN-LOG), Sure Independence Screening (SIS). For classification methods, we investigated 9 machine-learning classifiers: $L^2$-Logistic Regression ($L^2$-LOG), Kernel Support Vector Machines (KSVM), Linear-SVM (LSVM), Adaptive Boosting (AdaBoost), Random Forest (RF), Neural Network (Nnet), K-nearest neighborhood (KNN), linear discriminant analysis (LDA), Naive Bayes (NB). A total of 54 cross-combinations can be derived from the 6 feature selection methods and 9 classification methods. All conventional radiomics methods were implemented using R package (version 3.6.3, http://www.R-project.org). The following R packages were used for feature selection methods: 'SIS' (SIS), 'VSURF' (RF), 'LiblineaR' ($L^1$-LOG and $L^1$-SVM), 'Energy' (DC), 'glmnet' (EN-LOG), while the following packages were used for classification methods: 'e1071' (KSVM), 'LibLineaR' (LSVM and $L^2$-LOG), 'adabag' (AdaBoost), 'randomForest' (RF), 'nnet' (Nnet), 'knn' (KNN), 'MASS' (LDA), 'e1071' (NB).

We extracted handcrafted radiomics features for each patient, including 19 features from First Order Statistics (FOS), 24 features from Grey-Level Cooccurrence Matrix (GLCM), 16 features from Grey-Level Run Length Matrix (GLRLM), 16 features from Grey-Level Size Zone Matrix (GLSZM), 5 features from Neighboring Grey Tone Difference Matrix (NGTDM), and 16 features based on 3D shape of tumors. The 16 shape-based features were extracted from the segmentation mask, while the remaining $(19 + 24 + 16 + 16 + 5) = 80$ FOS and textural features (GLCM, GLRLM, GLSZM, and NGTDM) were extracted from the preprocessed PET and CT patches respectively. FOS and textural features also were recomputed after different wavelet decomposition in three directions (x, y, z) of PET and CT. Performing low-pass or high-pass wavelet filters along x, y, or z directions resulted in 8 decompositions of the original image (LLL, LLH, LHL, LHH, HHH, HLL, HHL, and HLH). Therefore, FOS and textural features were extracted from a total of $(1 + 8) = 9$ decompositions of PET and CT (including original ones). Consequently, the number of handcrafted radiomics features is $80 \times 9 \times 2 + 16 = 720 \times 2 + 16 = 1456$. A total of 720 features were extracted from PET or CT, while 16 features were extracted from tumor segmentation masks.

Redundant features with Spearman's correlation > 0.7 were first eliminated. Then, the remaining ones were fed into a feature selection model to screen for robust and relevant features. The number of the selected features derived from each selection method was regarded as a hyper-parameter and was individually optimized through grid searching from 1-30 with 1 as a step. Specifically, for each selection-classification combination, we tried to remain 1-30 features from the selection model and fed them into the classification model. The remaining number resulting in the highest internal validation result was chosen. Therefore, the number of the selected features for each combination is different. The average number of selected features derived from each selection method is as follows: $L^1$-LOG (10.14), $L^1$-SVM (8.53), SIS (6.43), DC (6.85), EN-LOG (8.26), and RF (4.57). Moreover, we regarded TNM stage as a clinical feature because it showed significant relevance to PFS in our univariate analysis (**Supplementary Table S1**). Both the selected radiomics features and clinical feature (TNM stage) were fed into a classification model for statistical analysis. The top-3 optimal combinations were identified based on the internal validation and then were further validated on the external cohort.

**Deep Learning-based Radiomics (DLR) Analysis**

We developed a multi-modality DLR model to directly predict 5-year PFS from pretreatment PET/CT images in an end-to-end manner. This DLR model is a 3D CNN and its architecture is based on 3D Deep Multi-modality Collaborative Learning (3DMCL) (29), a deep multi-modality architecture for predicting the distant metastases of soft-tissue sarcoma from PET/CT images. It takes as input a pair of preprocessed PET/CT patches (segmented ROIs), and outputs normalized probabilities for both classes (0 or 1). Note that although segmentation is not compulsory for DLR, we used the segmented ROIs in our DLR models for a fair comparison with conventional radiomics methods. The DLR network architecture is shown in **Figure 3**, which consists of two separate branches processing the PET and CT images in a simultaneous manner. For each PET and CT branch, there are five convolutional layers (Conv1-5) of 16, 32, 64, 128, and 256 filters with kernel sizes of 5, 3, 3, 3, and 3, respectively. The middle four convolutional layers (Conv2-5) are with a stride of 2 to reduce the resolution of feature maps. To fuse the feature maps derived from both branches, the output of PET and CT branches are concatenated together, and then are fed into another convolutional layer (Conv6) of 768 filters with kernel size of 3. Each convolutional layer is followed by a batch normalization layer and an Exponential Linear Unit (ELU) activation. The features maps obtained from the last convolutional layer (Conv6) are considered as deep features with high-level semantic information related to tumor characteristics. Finally, four fully-connected layers (FC1-4), which have 1024, 512, 256 and 2 nodes respectively, are added to perform survival prediction based on deep features. The first three fully-connected layers (FC1-3) are followed by a ReLU activation and a dropout layer, while the last layer (FC4) is followed by a softmax classifier layer which outputs normalized probabilities of two classes (0 or 1). TNM stage, as a high-level clinical feature, can be concatenated with the third fully-connected layer (FC3).

We first developed a DLR model taking PET, CT, and TNM stage as input (named PET+CT+TNM model). Then, we further established three degraded DLR models to evaluate the individual values of PET, CT, and TNM stage: (1) PET+CT model: a DLR model taking PET and CT as input. In this model, TNM stage was not concatenated with FC3 layer. (2) PET+TNM model: a DLR model taking PET and TNM stage as input, in which CT branch was truncated and only PET branch was connected to Conv6 layer. (3) CT+TNM model: a DLR model taking CT and TNM stage as input, in which PET branch was truncated and only CT branch was connected to Conv6 layer.

All DLR models were implemented using Keras with a Tensorflow backend on a 12GB TITAN V GPU. We used the Adam optimizer with a batch size of 32 and a learning rate of 0.0001 for training the model. Cross-entropy loss function was used as the loss and our training stops when there are no further drops in the total loss. All the hyper-parameters, including batch size, learning rate, dropout rate, and regularization terms, were chosen through 5-fold cross-validation within the internal cohort. During the training stage, data augmentation was applied to the input images in real-time to avoid overfitting. The used data augmentation techniques included random translations up to 8 pixels, random rotations up to 15 degrees, and random flipping along 3 axes. We sampled an equal number of positive and negative samples during the data augmentation process to minimize the problem introduced by unbalanced classes.

**Statistical Analysis**

For clinical and conventional PET parameters including age, gender, EBV status, histology, BMI, T stage, N stage, TNM stage, $SUV_{max}$, $SUV_{mean}$, MTV, and TLG, frequencies with percentages were used to describe categorical variables; medians or means with ranges were used to describe continuous characteristics. Differences of these parameters between the internal and external cohorts were assessed using the Mann-Whitney test (for continuous characteristics) and the $\chi^2$ test or Fisher's exact test (for discrete characteristics). Univariate and multivariate analyses were performed using Cox proportional hazard regression. Factors with $P < 0.1$ in univariate analysis were taken into multivariate analysis. Cox proportional regression analyses were implemented using SPSS (version 22.0; IBM Inc., New York, USA). $P < 0.05$ was considered statistically significant.

The performance of conventional radiomics methods and DLR models were evaluated using AUC and testing error. The statistical significance between AUCs was tested via DeLong's method using R packages (version 3.6.3, http://www.R-project.org).

Furthermore, survival analyses using Kaplan-Meier method were performed for risk group stratification. Specifically, patients with positive/negative prediction results were stratified into high/low-risk groups, and then a two-sided log-rank test was used to compare the two groups. The survival analyses were performed using SPSS.

## RESULTS

### Patients and Cox Regression Analyses for Clinical and Conventional PET Parameters

A total of 170 advanced NPC patients (female = 40, male = 130; median age = 46, range 16-78) were included in the internal cohort (**Table 1**). Among these patients, 121 patients (71.18%) were diagnosed with TNM stage III, and 49 patients (28.82%) were TNM stage IVa; 126 patients (74.12%) were non-keratinizing undifferentiated NPC; 80 patients (47.06%) were infected with EBV. Induction chemotherapy (IC) was given to 147 out of 170 (86.47%) patients, while concurrent chemoradiotherapy (CCRT) was used for 88 out of 170 (51.76%) patients. Furthermore, 77 out of 170 (45.29%) patients were treated with IC + CCRT. The mean $SUV_{max}$ value for primary tumor was 11.40 g/ml (range, 2.74-47.49 g/ml), the mean MTV value was 30.34 ml (range, 0.17-152.99 ml), and the mean TLG value was 155.92 g (range, 0.45-864.03 g). The detailed baseline data of the external cohort was also summarized in **Table 1**.

Furthermore, 48 out of 170 (28.24%) patients (for the internal cohort) and 45 out of 87 (51.72%) patients (for the external cohort) suffered disease progression or death after treatment within the first 5 years. Among clinical and conventional PET parameters, only TNM stage was significantly associated with PFS in univariate analysis ($P = 0.047$) on the internal cohort, while no factors showed a significant correlation with PFS on the external cohort (**Supplementary Table S1**). Furthermore, MTV and TLG showed significantly associated with PFS in univariate analysis ($P = 0.002$ and $P = 0.002$) on the internal cohort using a cutoff value of 39.80 ml and 198.68 g, respectively. However, none of these attributes showed a significant correlation with PFS in multivariate analysis on both cohorts (**Supplementary Table S2**).

### Prognostic Performance of Conventional Radiomics

**Figure 4A** shows that the combinations with top-five mean AUC are: RF + RF (0.796 ± 0.033), RF + AdaBoost (0.783 ± 0.041), SIS + LSVM (0.778 ± 0.024), $L^1$-LOG + $L^2$-LOG (0.772 ± 0.027), and $L^1$-LOG + KSVM (0.769 ± 0.038). As demonstrated in **Figure 4B**, the combinations with bottom-five mean testing error are: RF + RF (0.267 ± 0.037), RF + KSVM (0.283 ± 0.034), RF + AdaBoost (0.286 ± 0.029), $L^1$-LOG + KSVM (0.298 ± 0.026), and RF + KNN (0.301 ± 0.032).

Scatterplot was used to screen out the combinations with both high AUC and low testing error. **Figure 5** illustrates that the optimal combinations are RF + RF (AUC = 0.796 ± 0.033, testing error = 0.267 ± 0.037), RF + AdaBoost (AUC = 0.783 ± 0.041, testing error = 0.286 ± 0.029), and $L^1$-LOG +

KSVM (AUC = 0.769 ± 0.038, testing error = 0.298 ± 0.026), which show higher prognostic performance than other combinations.

**Comparison between DLR and Conventional Radiomics**

We first compared the performance of DLR models using ROC analysis. **Figure 6** and **Table 2** show that the PET+CT+TNM model has the highest prognostic performance in the internal validation (AUC = 0.842 ± 0.034, 95% CI: 0.801-0.889; testing error = 0.194 ± 0.029) and external validation (AUC = 0.823 ± 0.012, 95% CI: 0.787-0.862; testing error = 0.238 ± 0.008), while PET+CT model and PET+TNM model show lower but also good prognostic performance in the internal validation (AUC = 0.825 ± 0.041 and 0.818 ± 0.029, 95% CI: 0.775-0.870 and 0.762-0.862; testing error = 0.223 ± 0.035 and 0.218 ± 0.024) and external validation (AUC = 0.819 ± 0.017 and 0.796 ± 0.009, 95% CI: 0.778-0.856 and 0.747-0.829; testing error = 0.241 ± 0.009 and 0.262 ± 0.006). However, CT+TNM model shows much lower prognostic performance in the internal validation (AUC = 0.657 ± 0.055, 95% CI: 0.596-0.718; testing error = 0.375 ± 0.048) and external validation (AUC = 0.645 ± 0.021, 95% CI: 0.591-0.709; testing error = 0.403 ± 0.011).

Then, we further compared the optimal DLR model (PET+CT+TNM model) with the optimal conventional radiomics methods (RF + RF, RF + AdaBoost, and $L^1$-LOG + KSVM). **Figure 6** and **Table 3** show that the DLR model has significantly higher prognostic performance than the three conventional radiomics methods in the internal validation (AUC = 0.842 vs 0.796, 0.783, and 0.769; $P$ = 0.038, 0.046, and 0.027) and external validation (AUC = 0.823 vs 0.782, 0.767, and 0.755; $P$ = 0.043, 0.026, and 0.015).

**Survival Analysis for Risk Group Stratification**

We used the optimal clinical signature (TNM stage), conventional radiomic signature (RF + RF), and DLR signature (PET+CT+TNM model) for risk group stratification. **Figure 7** shows the Kaplan-Meier curves of the high- and low-risk patient groups stratified by the clinical signature (**A, D**), conventional radiomic signature (**B, E**), and DLR signature (**C, F**). The conventional radiomics signature and DLR signature enabled significant difference between the high- and low-risk patient groups in the both internal and external cohorts ($P$ < 0.001), while the clinical signature failed in the external cohort ($P$ = 0.042 and 0.177 for the internal and external cohorts).

**DISCUSSION**

In this study, we proposed an end-to-end multi-modality DLR model to directly predict 5-year PFS from pretreatment PET/CT images in advanced NPC patients. The main finding of this study is that multi-modality image-derived DLR outperformed clinical indicators, conventional PET predictors, and conventional radiomics methods.

Pretreatment medical images contain much more information than diagnosis and TNM stage. However, clinical oncologists usually only employ "visible" information in routine clinical practice and treatment planning. Conventional PET parameters, such as MTV or TLG, have been demonstrated to serve as an independent predictor for survival prediction (14, 15, 17). However, in this study, the conventional PET parameters ($SUV_{max}$, $SUV_{mean}$, MTV, and TLG) showed no significant association with PFS (Supplementary Table S1 and S2), which is consistent with some studies where the conventional PET parameters also failed to show significant association for survival prediction (40-42). Therefore, we suggest that the conventional PET parameters have limited reproductivity among different studies and different centers.

In radiomics analysis, we found that both conventional radiomics and DLR showed consistent prognostic performance in the internal and external cohorts, despite that the distribution of some clinical characteristics and progression rate showed significant differences (e. g., N2 and N1) between the two cohorts, which indicates strong generalizability. This is possibly because radiomics extracts features from PET and CT images, which employs intra-tumor information such as tumor texture, intensity, heterogeneity, and morphology (43). Intra-tumor information can reflect intra-tumor heterogeneity that is caused by the genetic instability and potentially leads to the drug resistance and treatment failure (44). The uptake of $^{18}$F-FDG in tumor cells can reflect the intra-tumor heterogeneity by exhibiting variations in glucose metabolism of different tumor regions (45). However, the conventional parameters derived from PET/CT (e.g., MTV and TLG) can only represent the apparent metabolism information, which thereby fail to reflect the intra-tumor heterogeneity (46).

Conventional radiomics depends heavily on handcrafted feature extraction and manual tuning of statistical models. Therefore, in many studies (21, 25, 47-49), different feature extraction schemes, feature selection methods, and statistical models were investigated to identify optimal conventional radiomics methods for a specific clinical target. Our conventional radiomics analysis followed the scheme proposed by Zhang et al. (21), and we derived 54 cross-combinations from 6 feature selection methods and 9 classification methods. Considering both AUC and testing error (Fig. 4), the optimal combinations were RF + RF and RF + AdaBoost, which is consistent with Zhang et al.'s study (21). DLR is considered as a replacement for conventional radiomics and has demonstrated superior prognostic performance in some studies (31-33). However, there are few studies where DLR was comprehensively compared with conventional radiomics. In most literature (31-34, 36-38), only a few conventional radiomics methods were included in the comparison, but there is no guarantee that the included methods are optimal for their clinical targets. In this study, the three optimal conventional radiomics methods chosen from 54 cross-combinations were chosen as the benchmarks for further comparison with our DLR model. Our DLR models showed the highest performance in prognosis (**Table 3**), and revealed the best risk group stratification for advanced NPC patients when compared to clinical and conventional radiomics signatures (**Figure 7**). Through the comprehensive comparison, we suggest that DLR is superior to conventional radiomics in the prediction of long-time PFS in advanced NPC. We attribute this superiority to three main reasons: First, DLR allows for automatic learning of robust and relevant features from multi-modality PET/CT images, which reduces the human bias (or even errors) caused by handcrafted feature extraction. Second, the automatic feature learning discovered high-level semantic features that may be overlooked by the manually-defined feature extraction. Third, it has been widely recognized that CNNs have the outstanding ability for pattern recognition (50), so DLR can inherit this ability by employing CNN architectures.

Although DLR has been applied for prognoses of NPC in several previous studies (35, 37, 38), there still exist the following limitations: Peng et al. (35) used 2.5D CNNs to extract deep features rather than to directly predict survival outcomes in an end-to-end manner, and the deep features were extracted from informative slices instead of from the entire tumor volume. Jing et al. (38) addressed the aforementioned limitations by employing a 3D end-to-end DLR model that directly predicts the risk of disease progression based on the entire tumor volume. Their study demonstrated that end-to-end DLR models are more effective to extract relevant features and showed higher prognostic performance, but their study was limited by only using single-modality MRI. Zhang et al. (37) also used an end-to-end DLR model to directly predict Distant Metastasis-Free Survival (DMFS), but their study suffered from the same limitations as Peng et al.'s and Jing et al.'s: they also used a 2D CNN that makes decisions based on maximum three tumor slices, and their study was also limited by using single-modality MRI. In this study, we have overcome these limitations by developing a 3D multi-modality end-to-end DLR model that was purposely optimized for PET/CT images. In contrast to Peng et al.'s study where deep features were separately extracted from PET and CT images, our DLR model can simultaneously extract complementary deep features from both PET and CT images. In the experiments, since Zhang et al.'s and our studies both focused on similar clinical targets (DMFS and PFS) and used the same evaluation metrics (AUC), their result can be directly compared with ours. Our multi-modality DLR model (PET+CT+TNM model) showed higher prognostic performance than Zhang et al.'s method (AUC of 0.842 versus 0.795), in spite of the fact that Zhang et al. employed additional information (e.g., EBV DNA and treatment regimens) for prediction. The result demonstrates that, by overcoming the aforementioned limitations with a 3D PET/CT-based end-to-end DLR model, our study indeed gained improvements over previous DLR studies in NPC.

TNM staging system is widely used in clinic for risk stratification and decision-making for treatment, but TNM stage alone is limited for survival prediction (8). In the survival analysis (**Figure 7**), no significant difference between the high- and low-risk patient groups was identified in the external cohort ($P = 0.177$), which indicate that TNM stage lacks accuracy in risk group stratification. Nevertheless, existing studies have shown that TNM stage combined with other clinical signatures may improve the performance for risk discrimination (51-53). Our results show that TNM stage was significantly associated with PFS in univariate analysis (Table 1). Therefore, we further combined TNM stage into our DLR models (PET+CT+TNM, PET+TNM, and CT+TNM models) to evaluate its capability for further improving the prognostic performance. For evaluating the value of TNM stage, we established a DLR model excluding TNM stage (PET+CT model). The results showed that the PET+CT+TNM model has improved prognostic performance compared to the PET+CT model (**Table 2**), which suggests that TNM stage can provide supplementary information and enhance the prognostic performance of survival prediction in NPC patients. We further compared the multi-modality DLR model (PET+CT+TNM model) with two single-modality DLR models (PET+TNM and CT+TNM model) for additional comparisons (**Table 2**). The results showed that the PET+CT+TNM models outperformed the PET+TNM and CT+TNM models, which suggests that integrating both metabolic (from PET) and anatomical (from CT) information helps achieve higher prognostic performance.

There are some limitations with the current study. Firstly, EBV status was determined by testing plasma anti-EBV IgA antibodies rather than plasma EBV DNA levels in our center, and EBV status was missing for 35.02% of the patients, which might limit the accuracy of statistical analysis. This was because the technology was only recently adopted in our center. Secondly, our internal cohort is relatively small (170 patients). To alleviate this limitation, we performed 5-fold cross-validation within the internal cohort for the establishment and internal validation of all prognostic methods. Then, all methods were further validated on the external cohort acquired from a different center. These settings help to avoid the sampling bias led by single random training-validation split and gain generalizable results in a small dataset (54). Thirdly, the progression rate of the external cohort is higher than that of internal cohort (51.72% vs. 28.24%). This is attributed to the fact that patients in the external cohort were enrolled from SPHIC, a young center established in June 2014, with the enrollment time from June 2014 to May 2019. Since we focused on the survival analysis of 5-year PFS, many patients without disease progression were excluded as they were not followed up for sufficient time (more than 5 years). In contrast, all patients with disease progression were enrolled, which inevitably increased the progression rate in the external cohort. Nevertheless, the variations between the internal and external cohorts did validate that radiomics methods have strong generalizability. Radiomics methods performed patient-wise prediction based on PET/CT of each individual patient, such that their prediction results were not affected by the progression rate in the cohorts. Fourthly, DLR models are limited by the 'black box' problem: an algorithm operates in an obscured space that is inaccessible to humans (55). Despite this, existing clinical studies (30-38), even conducted at a very large scale (56), demonstrate the advantages and feasibility of using these models for clinical purposes. A recent study has attempted to make advances into this problem and is leading toward a solution that physicians and patients can better understand how DLR makes decisions from medical images (57). Finally, since we focused on the comparison of DLR and conventional radiomics in their capability for survival prediction in advanced NPC, we, therefore, optimized our DLR models based on the existing reliable and proven deep learning architectures (3DMCL (29)). In our future work, we plan to evaluate other deep learning architectures and apply the proposed DLR model to other types of cancer diseases.

## CONCLUSION

We introduced and evaluated an end-to-end multi-modality DLR model in predicting 5-year PFS of patients with advanced NPC using pretreatment PET/CT images. Our results demonstrated that our DLR models improved prognostic performance over conventional radiomics methods. Furthermore, our DLR model can facilitate risk group stratification for advance NPC patients, where the DLR signature demonstrated greater accuracy for risk group stratification than the clinical signature and conventional radiomics signature. Our study suggested that multi-modality DLR derived from pretreatment PET/CT could provide complementary values to the current TNM staging, and guide the individual treatment practice in routine clinical care for NPC patients.

**DATA AVAILABILITY STATEMENT** The original contributions presented in the study are included in the article/Supplementary Material. Further inquiries can be directed to the corresponding author.

**ETHIC STATEMENT** All procedures involving human participants were carried out in accordance with the ethical standards of the institutional and/or national research committee and with the 1964 Helsinki Declaration and its later amendments or comparable ethical standards. Informed consent was obtained from all individual participants included in the study.

**AUTHOR CONTRIBUTIONS** SLS and DDF conceptualized and designed the study. BXG, MYM, LB, and JK performed analysis. BXG and MYM interpreted the data and drafted the manuscript. LB, JK and SLS revised the manuscript. All authors contributed to the article and approved the submitted version.

**FUNDING** This work was funded by National Natural Science Foundation of China (Grant number 81901778 and 81971648).

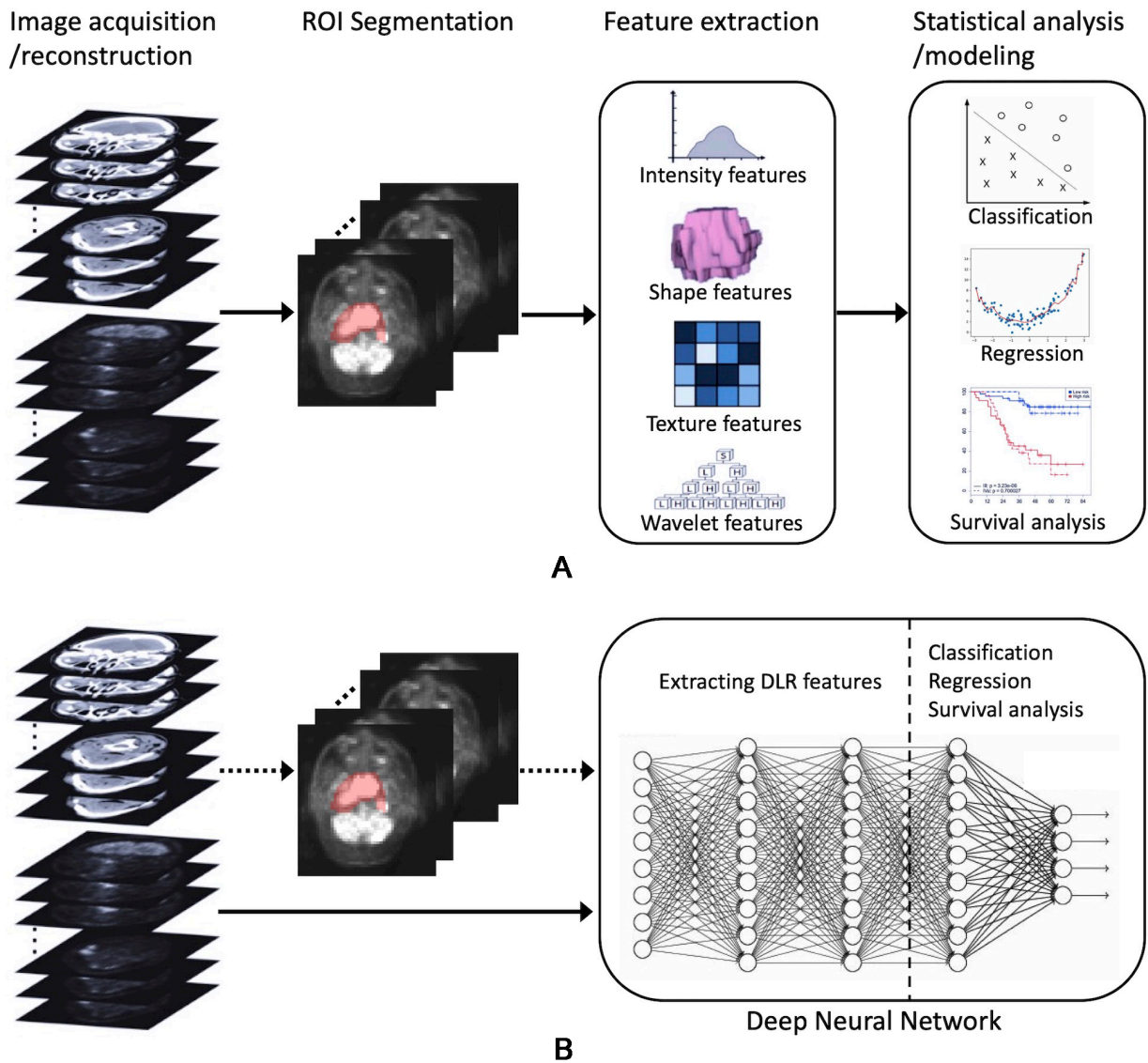

**FIGURE 1** | An illustration of radiomics process. (A) Conventional radiomics consisting of 4 steps: image acquisition/reconstruction, ROI segmentation, feature extraction, and statistical analysis/modeling. (B) DLR: its feature extraction and analysis are joint learned using a deep neural network. The dotted arrow indicates the optional ROI segmentation operation.

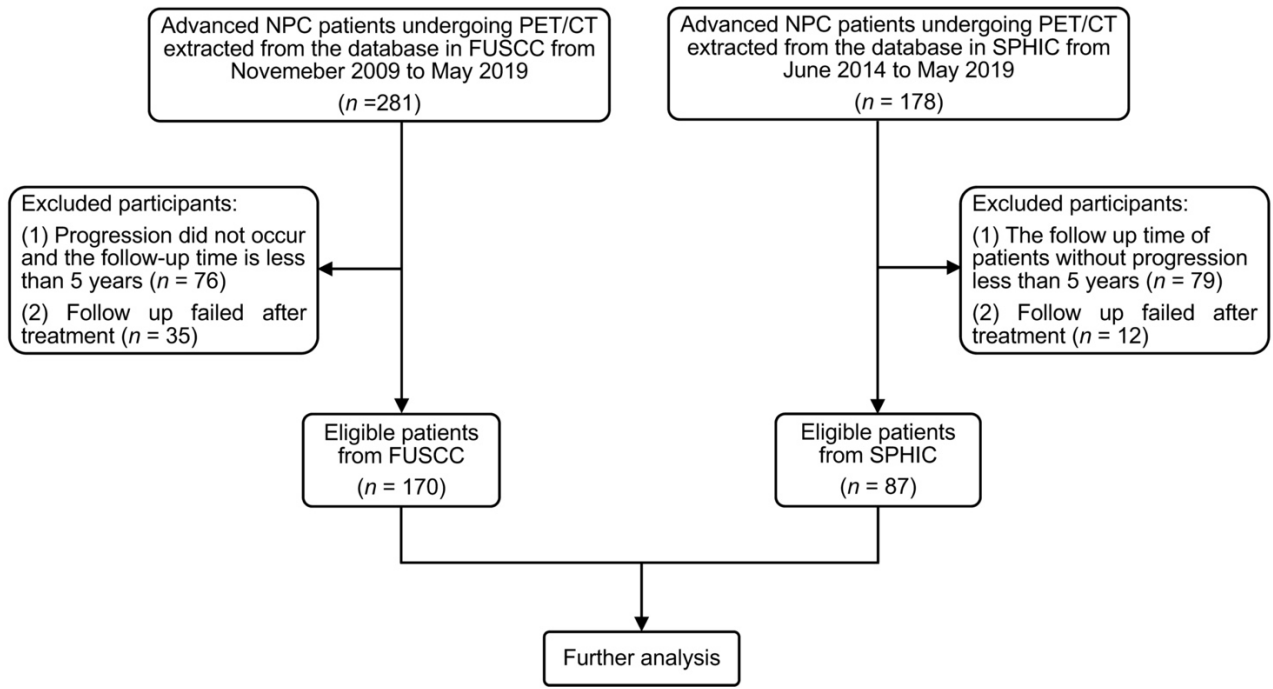

**FIGURE 2** | The flowchart of patient inclusion and exclusion. *NPC, Nasopharyngeal Carcinoma; FUSCC, Fudan University Shanghai Cancer Center; SPHIC, Shanghai Proton and Heavy Ion Center.*

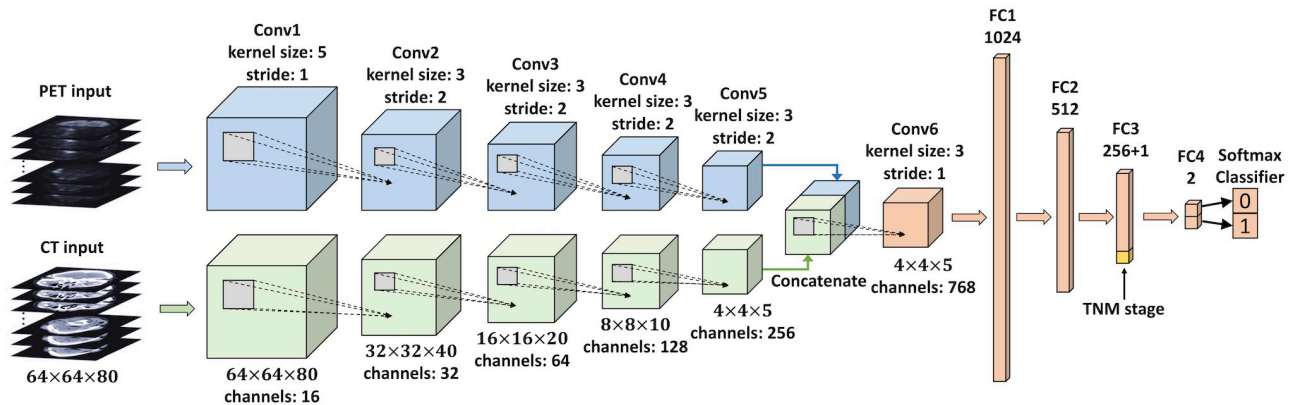

**FIGURE 3** | An illustration of the CNN used in the DLR analysis. This network takes a pair of preprocessed PET/CT patches (segmented ROIs) as input, and the final layer (FC4) outputs normalized probabilities for both classes (0 or 1). The clinical feature (TNM stage) is concatenated with FC3.

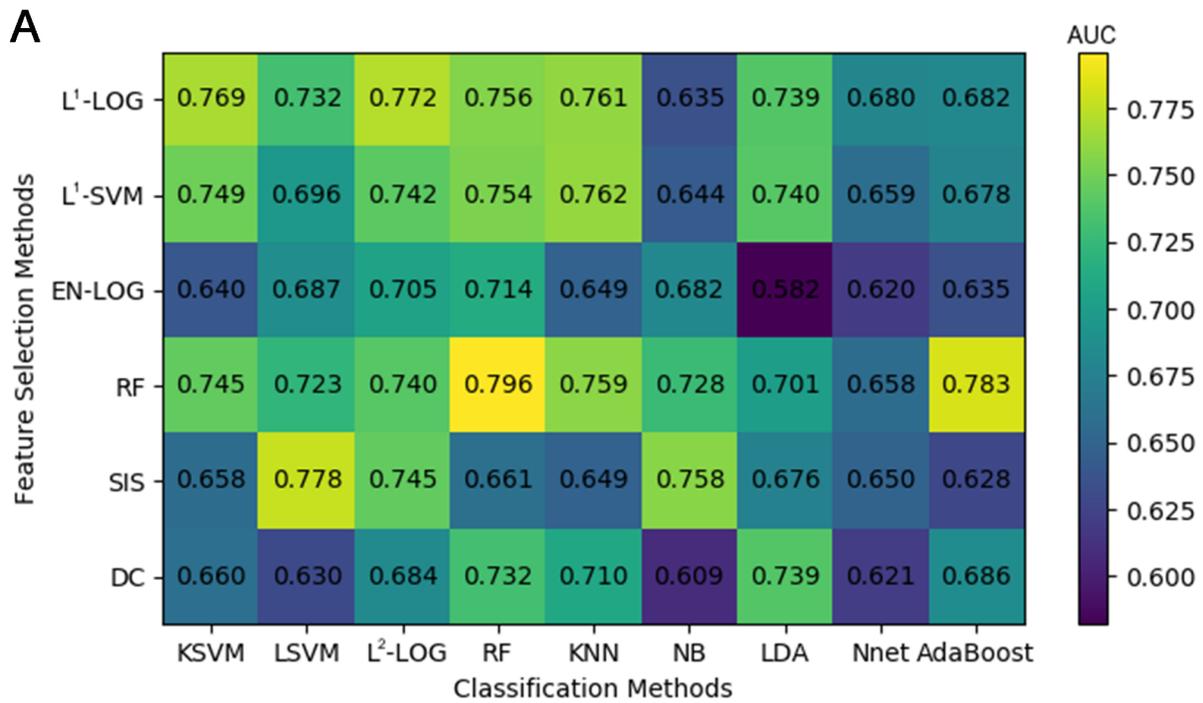
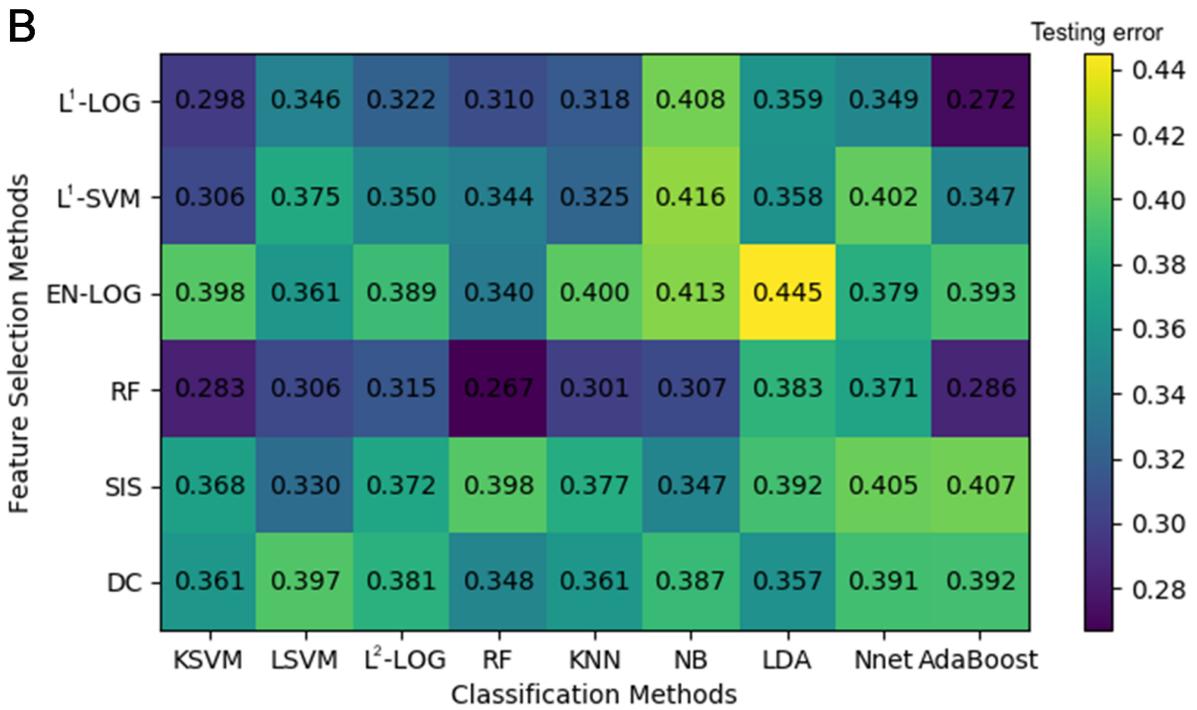

**FIGURE 4 |** A heatmap depicting the mean internal validation AUC (A) and testing error (B) of all 54 cross-combinations derived from 6 feature selection methods (in rows) and 9 classification methods (in columns).

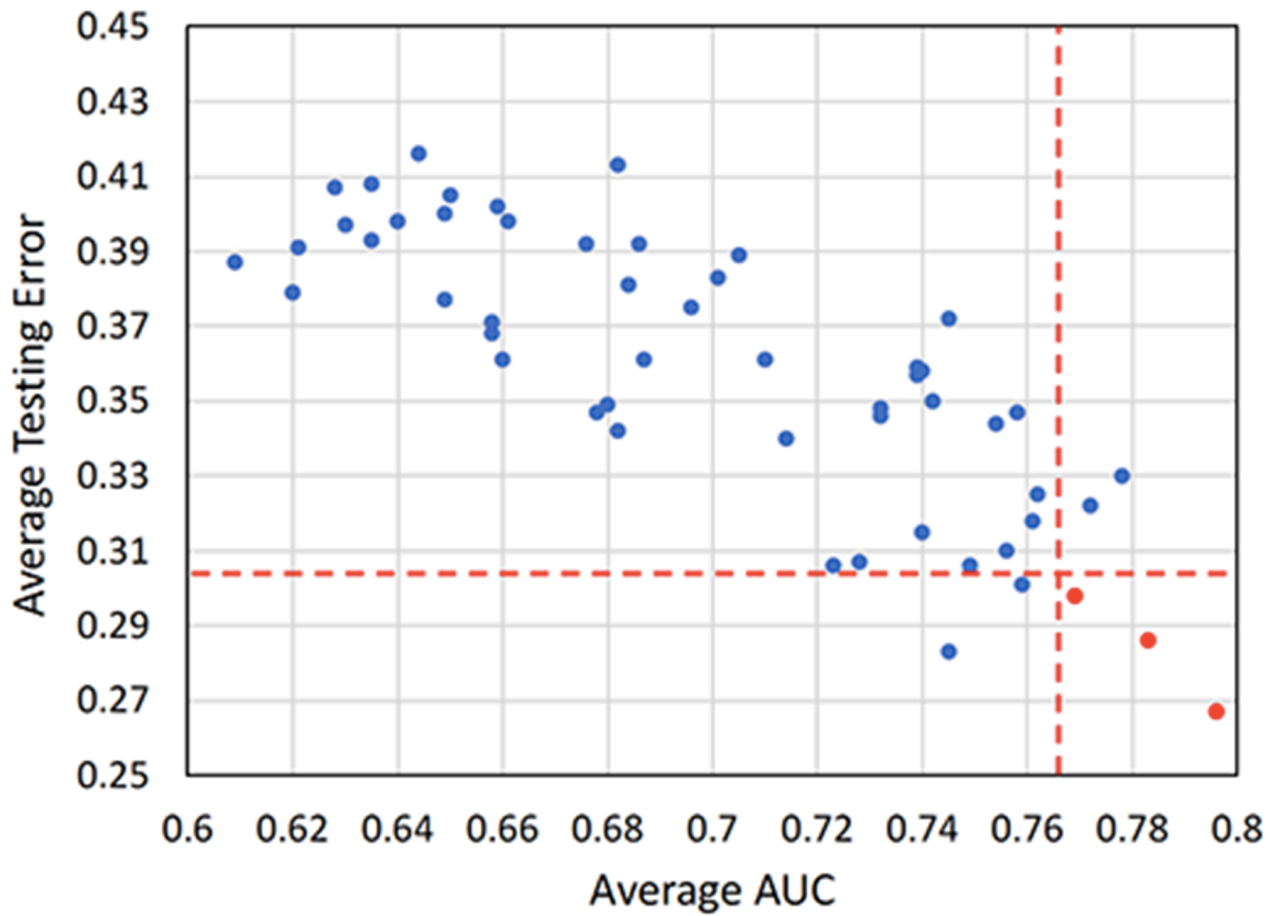

**FIGURE 5 |** A scatterplot depicting the mean internal validation AUC and testing error of all 54 cross-combinations derived from 6 feature selection methods and 9 classification methods. Two red dotted lines distinguish the combinations that show top-5 performance in AUC and in testing error. Three highly reliable and prognostic combinations that show top-5 performance in both AUC and testing error (RF + RF, RF + AdaBoost, and $L^1$-LOG + KSVM) are displayed in red points.

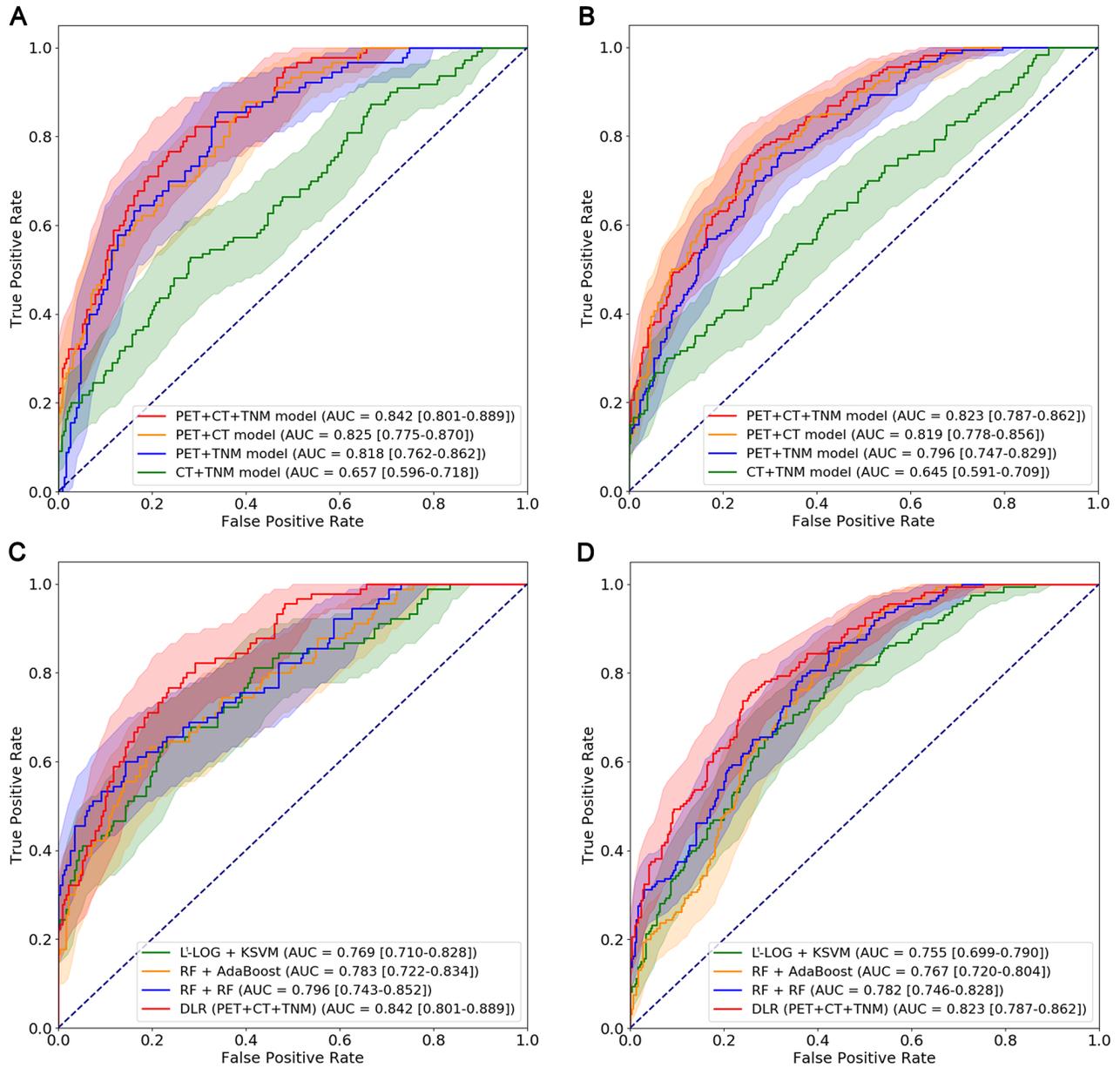

**FIGURE 6 |** ROC curves for comparison among four DLR models and comparison among optimal conventional radiomics methods and optimal DLR model on the internal cohort (A, C) and external cohort (B, D).

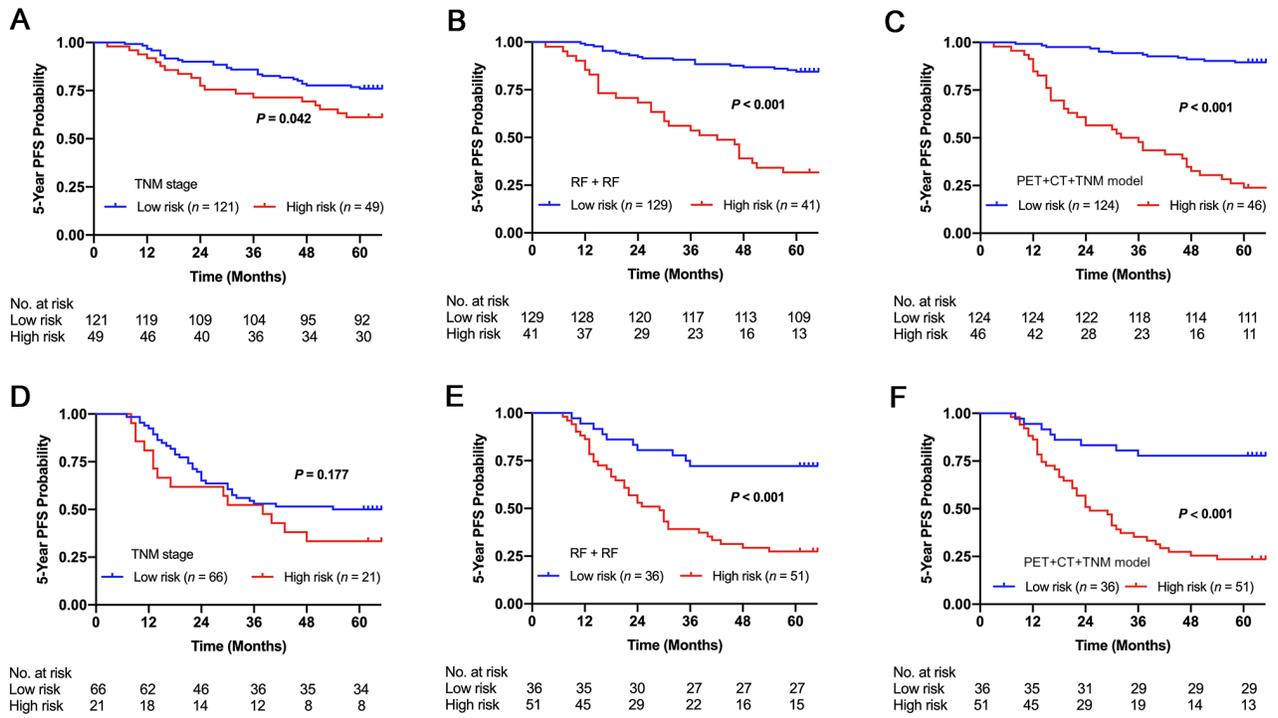

**FIGURE 7** | Kaplan-Meier curves of risk group stratification based on clinical signature (TNM stage), conventional radiomic signature (RF+RF), and DLR signature (PET+CT+TNM model) on the internal cohort (A, B, C) and external cohort (D, E, F).

**TABLE 1** | Patient characteristics in the internal and external cohorts.

| Characteristics | Internal cohort (n = 170) | External cohort (n = 87) | P value |
|---|---|---|---|
| Age (years), median (range) | 46 (16-78) | 46 (14-71) | 0.220 |
| Gender | | | 0.265 |
|   Male | 130 (76.47%) | 72 (82.76%) | |
|   Female | 40 (23.53%) | 15 (17.24%) | |
| EBV antibody | | | 0.008 |
|   Positive | 80 (47.06%) | 58 (66.67%) | |
|   Negative | 24 (14.12%) | 5 (5.74%) | |
|   Unknown | 66 (38.82%) | 24 (27.59%) | |
| Histology, WHO Type [a] | | | 0.001 |
|   I | 4 (2.35%) | 0 (0%) | |
|   II | 40 (23.53%) | 4 (4.60%) | |
|   III | 126 (74.12%) | 83 (95.40%) | |
| BMI (Kg/m$^2$), mean (range) | 22.79 (15.40-31.41) | 24.23 (16.41-34.41) | 0.001 |
| T stage | | | 0.001 |
|   T1 | 54 (31.77%) | 21 (24.14%) | |
|   T2 | 21 (12.35%) | 10 (11.49%) | |
|   T3 | 72 (42.35%) | 55 (63.22%) | |
|   T4 | 23 (13.53%) | 1 (1.15%) | |
| N stage | | | 0.196 |
|   N0 | 11 (6.47%) | 3 (3.44%) | |
|   N1 | 34 (20.00%) | 24 (27.59%) | |
| N2 | 96 (56.47%) | 40 (45.98%) | |
| N3 | 29 (17.06%) | 20 (22.99%) | |
| TNM stage | | | 0.462 |
|   III | 121 (71.18%) | 66 (75.86%) | |
|   IVa | 49 (28.82%) | 21 (24.14%) | |
| Concomitant systemic treatment with IMRT | | | |
|   None IC or CCRT | 12 (7.06%) | 2 (2.30%) | <0.001 |
|   IC alone | 70 (41.18%) | 18 (20.69%) | |
|   CCRT alone | 11 (6.47%) | 3 (3.45%) | |
|   IC + CCRT | 77 (45.29%) | 64 (73.56%) | |
| Targeted Therapy | 26 (15.29%) | 8 (9.20%) | 0.243 |
| PET Parameters, mean (range) | | | |
|   $SUV_{max}$ (g/ml) | 11.40 (2.74-47.49) | 14.95 (3.87-67.29) | 0.001 |
|   $SUV_{mean}$ (g/ml) | 4.71 (2.63-8.41) | 5.33 (2.92-17.70) | 0.016 |
|   MTV (ml) | 30.34 (0.17-152.99) | 18.58 (1.06-61.27) | 0.001 |
|   TLG (g) | 155.92 (0.45-864.03) | 114.38 (3.10-906.42) | 0.017 |

[a] WHO Type I = keratinizing, WHO Type II = non-keratinizing (differentiated), WHO Type III = non-keratinizing (undifferentiated)

*EBV, Epstein–Barr virus; WHO, World Health Organization; BMI, body mass index; IMRT, intensity-modulated radiation therapy; IC, induction chemotherapy; CCRT, concurrent chemoradiotherapy; SUV, standardized uptake value; MTV, metabolic tumor volume; TLG, total lesion glycolysis.*

**TABLE 2** | Comparison among four DLR models.

| Method | Internal validation | | External validation | |
| --- | --- | --- | --- | --- |
| | AUC | Testing error | AUC | Testing error |
| PET+CT+TNM model | **0.842 (0.034)** | **0.194 (0.029)** | **0.823 (0.012)** | **0.238 (0.008)** |
| PET+CT model | 0.825 (0.041) | 0.223 (0.035) | 0.819 (0.017) | 0.241 (0.009) |
| PET+TNM model | 0.818 (0.029) | 0.218 (0.024) | 0.796 (0.009) | 0.262 (0.006) |
| CT+TNM model | 0.657 (0.055) | 0.375 (0.048) | 0.645 (0.021) | 0.403 (0.011) |

Note: The average results are reported with standard deviation in the parenthesis. The best result in each column is in bold.

*DLR, deep learning-based Radiomics.*

**TABLE 3** | Comparison among the optimal conventional radiomics methods and DLR model.

| Method | Internal validation | | External validation | |
| --- | --- | --- | --- | --- |
| | AUC | Testing error | AUC | Testing error |
| $L^1$-LOG + KSVM | 0.769 (0.038) | 0.298 (0.026) | 0.755 (0.015) | 0.306 (0.007) |
| RF + AdaBoost | 0.783 (0.041) | 0.286 (0.029) | 0.767 (0.016) | 0.297 (0.008) |
| RF + RF | 0.796 (0.033) | 0.267 (0.037) | 0.782 (0.012) | 0.279 (0.010) |
| DLR (PET+CT+TNM) | **0.842 (0.034)** | **0.194 (0.029)** | **0.823 (0.012)** | **0.238 (0.008)** |

Note: The average results are reported with standard deviation in the parenthesis. The best result in each column is in bold.

*DLR, deep learning-based Radiomics.*